\RequirePackage{fix-cm}

\documentclass[numbook, envcountsect, envcountsame, envcountreset, runningheads, smallextended]{svjour3} 
\smartqed  
\usepackage{graphicx}
\journalname{}
\usepackage[utf8]{inputenc}
\usepackage{amsmath}
\usepackage{amsfonts}
\usepackage{amssymb}

\usepackage{appendix}
\usepackage{enumerate}
\usepackage{gensymb}

\newcommand{\R}{{\mathbb R}}

\newcommand{\Int}{I^{\circ}}

\newcommand{\sF}{ {\mathcal F} }
\newcommand{\sD}{ {\mathcal D} }
\newcommand{\sG}{ {\mathcal G} }

\newcommand{\half}{ {\scriptstyle{\frac{1}{2}} } }
\newcommand{\vv}{v}
\newcommand{\RN}{  \tilde{P}  }
\newcommand{\RW}{P}
\newcommand{\bb}{\beta}

\newcommand{\pchaf}{\Lambda}
\newcommand{\localtime}{\ell}

\newtheorem{assumption}[theorem]{Assumption}

\begin{document}

\title{Change of drift in one-dimensional diffusions}


\author{Sascha Desmettre         \and
        Gunther Leobacher        \and 
        L.C.G. Rogers
}


\institute{S. Desmettre (corresponding author) \at
              Institute of Financial Mathematics and Applied Number Theory, Johannes Kepler University Linz, Altenbergerstra\ss{}e  69, 4040 Linz, Austria, 
              \email{sascha.desmettre@jku.at} 
           \and
           G. Leobacher \at
           Department of Mathematics and Scientific Computing,   University of Graz, Heinrichstr. 36, 8010 Graz, Austria
            \and 
            L.C.G. Rogers \at Statistical Laboratory, University of Cambridge, Wilberforce Road, Cambridge CB3 0WB, United Kingdom
}

\date{\today}

\maketitle

\begin{abstract}
It is generally understood that a given one-dimensional diffusion may be
transformed by Cameron-Martin-Girsanov measure change into another
one-dimensional diffusion with the same volatility but a different drift. But
to achieve this we have to know that the change-of-measure local martingale
that we write down is a true martingale. We provide a complete characterisation
of when this happens.This enables us to discuss absence of arbitrage in a generalized
Heston model including the case where the Feller condition for the volatility
process is violated.


\end{abstract}

\section{Introduction}\label{S1}
 Our original goal in this paper was to understand how change of measure works in the celebrated stochastic volatility model of Heston \cite{Heston1993}. When this model is specified, if the growth rate of the asset is not equal to the riskless rate then we need to change measure to a pricing measure in which the growth rate {\em is} the riskless rate. The question then arises: `Can this be done?' The answer we found was `Not always'; and in cases where it cannot be done, then general results say that there must be arbitrage (in a suitable sense). 

\medskip

We then realised that the question is closely related to changing the given drift of a one-dimensional diffusion to a different drift, using change of measure. This uses the Cameron-Martin-Girsanov theorem, but as is well known this very general result cannot be applied without care, the main point being to decide whether the local martingale we write down to do the change of drift is actually a martingale. In general, this is hard to decide, but in the special case that concerns us, where the drift is again a function of the diffusion, we are able to derive {\em necessary and sufficient conditions} for the change of measure to `work'; we present this in Section \ref{S2}, as an algorithm to be followed to decide for any particular situation, and we illustrate this with two interesting examples.
 It might appear that the result we give is a reprise of the main results of Mijatovi\'{c} \& Urusov \cite{MU2012b,MU2012a}, but what we do here is actually rather different. A standing assumption throughout \cite{MU2012b,MU2012a} is that if a boundary point of the diffusion can be reached in finite time, then the diffusion stops there. For the application we have in mind, in Section \ref{S3}, this assumption does not hold; we want to consider CIR processes which reach zero and immediately return. Unusually, this behaviour can be completely specified by an SDE with nice coefficients, but  more generally a diffusion which can reflect from a boundary point cannot be specified by an SDE without explicitly involving a local time term, as in the Tanaka SDE for reflecting Brownian motion. While solutions of SDEs are very general regular one-dimensional diffusions, they are not the most general examples; as is well known, the most general regular one-dimensional diffusion is specified by its scale function and speed measure, and for our purposes it is necessary to work in this setting. This requires us to identify the Markov-process form of the Cameron-Martin-Girsanov change-of-measure local martingale, and to understand its effect on the generator of the diffusion. All this is explained in Section \ref{S2};  we are discussing here the transformation of Markov processes by multiplicative functionals, a topic which goes back a very long way, to It\^o \cite{Ito1965}, Dynkin \cite{Dynkin} and references therein, and which is still of interest nowadays, see e.g.~Palmowski \& Rolski \cite{PR2002}, or \c{C}etin \cite{Ce2018}.

Our work shares common features with Hulley \& Platen \cite{HP2008,HP2011} and Kotani
\cite{Kotani2006}, who also work with the scale and speed
representation (the main result of \cite{Kotani2006} is also obtained
by Delbaen \& Shirakawa \cite{DS2002} using different methods). 
The question they answer is: If a
one-dimensional diffusion $X$ is in natural scale, when is it a martingale, and
not a local martingale? Our study determines when the change-of-measure local
martingale $Z$ is a true martingale, which includes the problem of Hulley \& Platen
\cite{HP2008,HP2011} and Kotani \cite{Kotani2006} as the special case $Z = X$.

\medskip

In Section \ref{S3} we turn to the Heston model for the stock price $S$ and the volatility $\vv$, which defines their evolution in the `real-world' probability $\RW$ as 
\begin{eqnarray}
dS_t/S_t &=& \mu(\vv_t)\,dt + \sqrt{\vv_t}\,(\,\rho dW_t + 
\rho' dW'_t\,),
\label{dS1}
\\
d\vv_t &=& \kappa(\theta-\vv_t) \, dt + \sigma \sqrt{\vv_t}\, dW_t.
\label{dvv}
\end{eqnarray}
Here, $W$ and $W'$ are independent Brownian motions, 
 $\kappa$, $\theta$ and $\sigma$ are strictly positive constants, and $\rho \in (-1,1)$ is the constant correlation between the Brownian motions driving stock and volatility. We write $\rho' := \sqrt{1-\rho^2}$. The function $\mu$ is continuous. In Heston's original paper and many other studies, $\mu$ is taken to be constant, as is the riskless rate of interest $r$.  Here we will take $r=0$ throughout in order to simplify notation; this loses no generality, as we could equally consider $S$ defined by \eqref{dS1} to be the discounted stock $e^{-rt}S_t$.

In option pricing papers on Heston's stochastic volatility model, it is typically assumed that a risk-neutral measure $\RN$ exists and that the dynamics are stated in the corresponding risk-neutral form; see, for example, the extensive textbook Rouah \cite{RO2013} and the references therein. 
Yet, the question of existence of such a risk-neutral measure is rarely
investigated -- save for the trivial case $\mu= r$. But absence of such a
risk-neutral measure implies existence of free lunch with vanishing risk, that is,
a form of arbitrage! A notable exception, where this problem is addressed, is
Wong \& Heyde \cite{WH2006}, where the authors give a solution to this problem assuming
the Feller condition, a condition which keeps the volatility process strictly positive.

However, the Feller condition is frequently violated 
in practice as has been  pointed out  in Albrecher et al.~\cite{MSWT2007} or Clark \cite{C2011} (consult in particular Table~6.3). Building on results in Mijatovi\'{c} \& Urusov 
\cite{MU2012a,MU2012b}, 
this problem is addressed for several stochastic volatility
models in Bernard,  Cui, \& McLeish \cite{BCM2017}, including the classical Heston model,
by modifying the model so that the volatility process is stopped as 
soon as it hits $0$. While this solves the problems incurred by a
violated Feller condition mathematically, this approach is not 
convincing from an economic point of view.

\medskip

In Section \ref{S3}, we show that in the classical Heston model where the function $\mu$ is constant, then failure of the Feller condition implies that there is no risk-neutral measure. However, if the drift $\mu$ is not constant, but satisfies a simple integrability condition at 0, we show that there is an equivalent local martingale measure (ELMM), still in the case where the Feller condition is not satisfied.  When the Feller condition {\em is} satisfied, we show that there is always an ELMM.

\medskip 
In the Appendix, as a gentle amusement, we directly construct a free lunch with
vanishing risk (FLVR), from which if follows by the celebrated Fundamental
Theorem of Asset Pricing (FTAP) of Delbaen \& Schachermayer \cite{DS1994,DS1998} 
that there is no
equivalent $\sigma$-martingale measure and {\it a fortiori} no ELMM.  This is a
rare application of the FTAP!

\medskip
Does it really matter if the Feller condition fails, so that there is no ELMM? It does not; all that has happened is that we started off from a bad place, and what we should do is to immediately put ourselves into the risk-neutral measure (in effect, assume that $\mu = 0$). We gain nothing by being overly introspective about the growth rate of a stock, about which we know next to nothing in any case.

\section{Changing measure in a one-dimensional diffusion.}\label{S2}

We are going to begin with a regular diffusion taking values in an interval $ I \subseteq \R$. The killing measure is assumed to be zero.
We write $\Int$ for the interior of $I$, which could be equal to $I$. We also
set $a = \inf I$, $b=\sup I$, the endpoints of $I$.  The interval $I$ may be
the whole real line, it may contain endpoints or not. We write $C_0$ for the
space of continuous functions $f \colon I \rightarrow \R$   with limits at the
endpoints.


We let $\Omega = C(\R_+; I)$ be the  canonical path space with the canonical process $X_t(\omega) = \omega(t)$ and the raw filtration
 $\sF^{0}_t = \sigma(X_s: s \le t)$. If $\RW$ is the law of $X$ on $(\Omega, \sF^{0}_\infty)$ we let $(\sF_t)$ be the  universal completion
 of $(\sF^{0}_t)$.  We write $\sF= \sF_{\infty}$ for brevity. 
 We write $s$ for the scale function of $X$, and $m$ for its speed measure, so that the infinitesimal generator of $X$ is
 \begin{equation}
 \sG = \frac{1}{2}\; \frac{d^2}{dm\,ds}\; ;
 \label{Gdef}
 \end{equation}
 see, for example, Theorem VII.3.12 of Revuz \& Yor \cite{RY2004} (while noting the  different scaling factor for the speed measure there). If a boundary point is in $I$ ($a=0 \in I$, to fix ideas), there is  a boundary condition 
 \begin{equation}
 \frac{df}{ds}(0+) = 2 m(\{0\}) \, \sG f(0).
 \label{bc}
 \end{equation}
 See Proposition VII.3.13 of \cite{RY2004}, again noting the different scaling
factor here. We omit discussion of the situation $m(\{0\}) = \infty$,
corresponding to absorption at the boundary, as this is a special
case already dealt with in the earlier works, c.f.~Mijatovi\'{c} \& Urusov \cite{MU2012a,MU2012b}. 
Moreover, for volatility models, which are our main application, 
absorbing boundaries are not reasonable from an economic point of view.

Specifying the domain of functions on which $\sG$ acts is important. We fix some reference point $\xi \in \Int$ and define the domain $\sD$ of $\sG$ to be the set of all $f$  which are represented as
 \begin{equation}
 f(x) = c_1 + \int_\xi^x\Big( c_2 + \int_\xi^w 2 g(u) \, m(du)\Big) s(dw) 
 \label{dom}
 \end{equation}
for some constants $c_1, c_2\in \R$, and a function $g \in C_0$. In the case of boundary points in $I$,
$c_1,c_2$ have to satisfy the necessary boundary conditions \eqref{bc}. For example,
if $a=0 \in I$, then we could take $\xi = 0$ in \eqref{dom} and would then have
that $c_2=0$ in order to match the boundary condition there.  If $f \in \sD$ is
given as in \eqref{dom},  then it is immediate from \eqref{Gdef} that $\sG f =
g$, and it can be shown that \begin{equation}
f(X_t) - \int_0^t \sG f(X_u) \; du \quad \hbox{\rm is a local martingale.}
\label{flocmart}
\end{equation} 
 Now fix some measurable $h:I \rightarrow \R^+$ satisfying for all $c<d\in I^\circ$
that
\begin{equation}\label{eq:equivalent-speed}
\int_{[c,d]}h(x)\, m(dx)<\infty
\end{equation}
 and define 
 \begin{equation*}
 A_t = \int_0^t h(X_u) \; du.
\end{equation*}
 
We write $H_z := \inf\{ t>0: X_t = z \}$. 

The most common use of the first point of the next proposition is when $h
\equiv 1$ and we characterise the exact conditions under which the boundary
point is reached in finite time; see, for example, Theorem V.51.2 of
Rogers \& Williams \cite{RW2}. 

\begin{proposition}\label{prop1}
 Suppose that the endpoint $a$ is accessible: $s(a) > -\infty$. Then:
 \begin{enumerate}
 \item The following are equivalent:
\begin{itemize}
 \item[(i)]  $ \int_{a+} (s(x) - s(a))\, h(x) \; m(dx) < \infty $;
 \item[(ii)] $\RW[A_{H_a -} <\infty] >0. $
 \end{itemize} 
 \item The following are equivalent:
\begin{itemize}
 \item[(i)]  $ \int_{a+}  h(x) \; m(dx) < \infty $;
 \item[(ii)] $\RW^a[A_{H_a +} <\infty] = 1 $;
 \item[(iii)] $\RW^a[A_{H_a +} <\infty] >0. $
 \end{itemize}
 \end{enumerate}
 \end{proposition}

\begin{proof}
Writing $Y_t := s(X_t)$, we have that $Y$ is a diffusion in natural scale with speed measure $m^Y$ defined by
\begin{equation*}
m^Y\!(s(x)) = m(x).
\end{equation*}
The additive functional $A$ is thus expressed equivalently as
\begin{equation*}
A_t = \int_0^t (h \circ s^{-1})(Y_u) \; du.
\end{equation*}
Since $Y$ is in natural scale with speed $m^Y\!$, it can be represented as
\begin{equation*}
Y_u = B_{\tau_u}, \qquad   \tau_u = \inf\{ t: \pchaf{}_t >u\},\qquad
\pchaf{}^Y_t = \int_{s(I)} \localtime(t,y)\; m^Y\!(dy),
\end{equation*}
where $\localtime$ is the local time of the Brownian motion $B$, see Rogers \& Williams \cite[Theorem V.74.1]{RW2}. 

For Point 1 first assume that for all $c,d\in I^\circ$ we have  
$0<\int_I h(u)m(du)$. Then we construct another diffusion $Z$ with speed 
measure $m^Z$ defined by 
\[
m^Z(C):=\int_C (h\circ s^{-1})(y)dm^Y\!(y) \quad\text{for all measurable $C\subseteq s(I)^\circ$}\,,
\]
which is a regular diffusion by \cite[Remark (ii) after (V.47.5)]{RW2} and by
our assumption \eqref{eq:equivalent-speed} on $h$,
and
\[
Z_u=B_{\tau^Z_u}\,,\quad \tau^Z_u=\inf\{t\colon \pchaf^Z_t>u\}\,,\quad \pchaf^Z_t=\int_{s(I)} \localtime(t,y)m^Z(dy)\,.
\]
Without loss of generality we may assume $s(a)=0$, and we
write
 \[T_0=\inf\{t\colon B_t=0\}, \;H^Y_0=\inf\{t\colon Y_t=0\}\, (=H_a),\;
H^Z_0=\inf\{t\colon Z_t=0\}\,,\] and we note that
$\tau^Y_{H^Y_0}=T_0=\tau^Z_{H^Z_0}$.  The appropriate occupation time formula
\cite[eqn.(V.49.2)]{RW2} also holds for positive measurable 
(instead of bounded measurable) functions, by monotone convergence.  
Therefore
\begin{align*}
\int_0^{u}h(X_s)ds&=\int_{s(I)} \localtime(\tau_u,y)(h\circ s^{-1})(y)m^Y\!(dy)
=\int_{s(I)} \localtime(\tau_u,z)m^Z(dz)=\pchaf^Z_{\tau_u}\,,
\end{align*}
such that 
\(\pchaf^Z_{t}=\int_0^{\pchaf^Y_t}h(X_s)ds\,\) for all $t<T_0$ (where $\pchaf^Y$ and $\pchaf^Z$ are strictly increasing), and therefore
\[
A_{H^X_a-}=A_{H^Y_0-}=\int_0^{H^Y_0-}h(X_s)ds=H^Z_0\,.
\]

Point 1 now follows immediately from Theorem V.51.2 of \cite{RW2} applied 
to the diffusion $Z$ . 

Next consider the general case where $\int_{[c,d]} h(u)m(du)$ may 
vanish. We choose a positive function $f\colon I\to \R$
with $\int_{a+}f(u)m(du)<\infty$. 
By the first part, $\RW[\int_0^{H_a-} f(X_s)ds<\infty]>0$,
so 
\[
\RW[A_{H_a-}<\infty]>0
\Leftrightarrow \RW\Big[\int_0^{H_a-} (h+f)(X_s)ds<\infty\Big]>0\,.
\]
On the other hand, by our assumption on $f$,
\[
 \int_{a+} \big(s(x) - s(a)\big)\, h(x) \; m(dx) < \infty 
\Leftrightarrow 
 \int_{a+} \big(s(x) - s(a)\big)\, (h+f)(x) \; m(dx) < \infty \,.
\]
Thus Point 1 for the general case follows from the special case.

\medskip\noindent

2. If we start $Y$ at the boundary point $0$ and run til the first time $T^Y_1$
at which it reaches 1, then $\tau^Y_{T^Y_1}=T_1:=\{t\colon B_t=1\}$ and some
simple calculus gives us \begin{eqnarray*}
A_{T^Y_1} &=& \int_{0}^{T^Y_1} (h \circ s^{-1})(Y_u) \; du
\\
&=& \int_0^{T_1} (h \circ s^{-1})(B_v ) \; d\pchaf{}_v
\\
&=& \int_{s(I)} (h \circ s^{-1})(y )\, \localtime(T_1,y)\,  m^Y\!(dy)
\\
&=& \int_{I} h(x)\, \localtime\big(T_1, s(x)\big)\; m(dx).
\end{eqnarray*}

By the Ray-Knight theorem  Rogers \& Williams \cite[Theorem VI.52.1]{RW2} the process
$y \mapsto \localtime(T_1,1-y)$ is a BESQ(2) diffusion started
at $0$. So $E[ \localtime(T_1,1-y) ]$ is finite, continuous in $y$, 
positive for $y>0$. Thus
almost surely, $\localtime(T_1,1-y)$ is bounded for $y$ in $[0,1]$; 
hence if $\int_{a+} h(x) m(dx) <
\infty$ it follows that $A_{H_a+}$ is a.s finite -- part (ii) of the statement -- and
this implies part (iii) a fortiori. 
Going from part (iii) to part (i), if it
were the case that $\int_{a+} h(x) m(dx) = \infty$, then since $\localtime(T_1,s(x))$ 
is a.s.
bounded away from zero in a neighbourhood of 0, it has to be that $A_{H_a+}$ 
is a.s.~infinite; a contradiction.  \hfill$\square $
\end{proof}

\begin{remark}\label{rem:prop1}
 Of course, there is an analogous statement for an accessible upper boundary point.
\end{remark} 
Now suppose that $Z$ is a non-negative continuous local martingale, $Z_0=1$. 
 {\em Provided} $Z$ is a martingale, we can define a new probability 
$\tilde P$ on $\sF_\infty$ by the recipe
\begin{equation}
\frac{d\tilde P}{dP}\biggl\vert_{\sF_t} = Z_t\,,
\label{com}
\end{equation}
since by the martingale property  \eqref{com} implicitly defines a family 
of consistent finite-dimensional distributions which can be extended using the
Daniell-Kolmogoroff extension theorem, see Rogers \& Williams \cite{RW1}.
 To determine whether or not $Z$ is a martingale, define the stopping times 
 \begin{equation*}
 T_n := \inf\{t: Z_t >n \}, \qquad n  =2, 3, \ldots
 \end{equation*}
which reduce $Z$, and notice that it is possible to define, for every positive integer $n$, a probability $\RN_n$ on $\sF_{T_n}$ by 
\begin{equation*}
\frac{d\RN_n}{d\RW}\biggl\vert_{\sF_{T_n}} = Z_{T_n}.
\end{equation*}
But does the sequence $\big(\RN_n)_n$ extend to a probability measure $\RN$ the whole of $\sF$? The answer is in this simple result (see Stroock \& Varadhan \cite{SV}, Theorem 1.3.5), whose proof we give for completeness.

\begin{theorem}\label{thm1}
The local martingale $Z$ is a martingale if and only if for each $t>0$
\begin{equation}
\RN_n(T_n \leq t) \rightarrow 0 \qquad(n \rightarrow \infty).
\label{eq10}
\end{equation}
\end{theorem}

\begin{proof}
We have
\begin{eqnarray*}
1 &=& E[Z_{t \wedge T_n}]
\\
&=& E[ Z_t1_{ t < T_n} ] + E[Z_{T_n}1_{T_n \leq t}]
\\
&=& E[ Z_t 1_{t < T_n }] + \RN_n[ T_n \leq t].
\end{eqnarray*}
By Monotone Convergence, the first term on the right converges to $E[Z_t]$, so condition \eqref{eq10} is equivalent to the statement that $E[Z_t]=1$ for all $t>0$, which is the condition that $Z$ is a martingale.
\hfill$\square$
\end{proof}
Note that, in the case when $Z$ is a martingale, 
$\RN_n=\RN$ on $\sF_{T_n}$, for every $n$.\\

 When is the condition \eqref{eq10} for $Z$ to be a martingale satisfied? To answer this, we define
the ``reverse measure transformation''
\begin{equation}
\tilde{Z}_t := \frac{1}{Z_t},
\label{tildeZdef}
\end{equation}
which is a positive $\RN$-local martingale if $Z$ is a martingale.
Obviously, $T_n = \inf\{ t:\tilde{Z}_t < n^{-1} \}$. According to Theorem 2.3 \[ \hbox{\rm condition (2.7)} \iff \hbox{\rm $Z$ is a martingale} \iff \hbox{\rm $\tilde{P}$ is well defined}. \] The goal now is to determine when condition (2.7) is satisfied.

\medskip
We need to be more specific about the local martingales $Z$ that we consider.
If the diffusion $X$ was specified as the solution of an SDE
\begin{equation}
dX_t = \sigma(X_t) \, dW_t + \bb(X_t)\, dt, \qquad  X_0 = x_0,
\label{dXbis}
\end{equation}
with a pathwise-unique strong solution and $C^1$ coefficients $\sigma >0$, $\bb$, then we  consider  local martingales $Z$ of the form
\begin{equation}
dZ_t = c(X_t) Z_t \, dW_t, \qquad Z_0 = 1,
\label{Zsde}
\end{equation}
where $c$ is assumed $C^1$ for convenience. The SDE \eqref{Zsde} has the solution
\begin{eqnarray}
Z_t &=& \exp\biggl( \int_0^t c(X_u) \; dW_u - \half \int_0^t c(X_u)^2 \; du \biggr)
\label{Z_CMG}
\\
&=& {\varphi(X_t)}\;\exp \bigg( - \int_0^t \frac{\sG \varphi}{\varphi}(X_u)
\; du \biggr),
\label{Z_markov}
\end{eqnarray}
where $\sG$ is the generator of $X$, and
\begin{equation}
\log \big(\varphi(x)\big) = \int_{x_0}^x \frac{c}{\sigma}(y) \; dy.
\label{logphi}
\end{equation}
The equivalence of \eqref{Z_CMG} and \eqref{Z_markov} is a simple exercise with
It\^o's formula, and is beside the point. The point is that the form \eqref{Z_CMG} of $Z$ {\em requires} that the diffusion $X$ is specified as the solution of an SDE, the form \eqref{Z_markov} {\em does not}. So we shall proceed to assume that 
$Z$ has the form \eqref{Z_markov} for some strictly positive function $\varphi \in \sD$  which satisfies $\varphi(x_0)=1$. 
In this generality, it may happen that $\varphi$ vanishes in an
endpoint $a$ of $I$. In that case the integral in \eqref{Z_markov} might
diverge, but since $Z$ is a non-negative local martingale, and therefore a
supermartingale, the limit $Z_{H_a-}:= \lim_{t\to H_a-}Z(t)$ exists and we
may set $Z_{t}:= Z_{H_a-}$ while $X_t$ remains in $a$.

The process $Z$ defined by \eqref{Z_markov} is still a local martingale,
since using partial integration on \eqref{Z_markov} gives
\begin{align*}
dZ_t
&=\Big(d\varphi(X_t)-{\sG \varphi}(X_t) \, dt\Big)
\exp \bigg( - \int_0^t \frac{\sG \varphi}{\varphi}(X_u)
\; du \biggr)\,,
\end{align*}
and $d\varphi(X_t)-{\sG \varphi}(X_t) \, dt$ is the differential of a 
local martingale by \eqref{flocmart}.

The next question is how the change of measure \eqref{Z_markov} (if it {\em is} a
change of measure) transforms the diffusion $X$. To answer this, we let $\tilde\sD$ be the set of all functions $f$ such that $f\varphi \in \sD$.

 Then using  It\^o's formula 
it is a simple exercise to show that that for any $f \in \tilde\sD$
\begin{equation*}
Z_t\, \biggl(
f(X_t) - \int_0^t \tilde{\sG}f(X_u) \; du
\biggr)
\quad \hbox{\rm is a local martingale,}
\end{equation*}
where
\begin{equation}
\tilde{\sG}f = \frac{1}{\varphi} \bigl( \sG(f\varphi)-f\sG\varphi\bigr)\,.
\label{Gtilde_def}
\end{equation}
The following result relates the form of $\tilde{\sG}$ just found to the form \eqref{Gdef}.

\begin{proposition}\label{propgen}
We have
\begin{equation*}
\tilde{\sG} = \frac{1}{2}\; \frac{d^2}{d\tilde{m}\,d\tilde{s} }\,,
\end{equation*}
where $\tilde s$ and $\tilde m$ take the  simple forms
\begin{equation}
d\tilde{m} = \varphi^2 dm, \qquad d\tilde{s} = \varphi^{-2} ds.
\label{tilde_ms}
\end{equation}
\end{proposition}

\begin{proof}
Take some continuous finite-variation test function $\psi: I \rightarrow \R$
which vanishes off some compact set. In what follows, we shall assume that $I$
is open, so that there are no boundary conditions to deal with, and leave the
checking of what happens in the other cases to the reader. Using
integration-by-parts, we develop 
\begin{align*}
\lefteqn{\int 2 \psi(x) \varphi(x) \tilde{\sG}f(x) \; m(dx)}\\
&= \int 2\psi\, \big(\sG(f\varphi) - f \sG\varphi\big)\; dm
= \int \psi\,\biggl( \frac{d}{dm} \biggl( \frac{d}{ds}(f\varphi)\biggr) -f
\frac{d^2}{dm\,ds}\varphi\biggr) \; dm
\\
&= -\int \frac{d}{ds}(f\varphi) \; d\psi + \int \frac{d\varphi}{ds} \; d(\psi f)
= -\int \varphi \frac{df}{ds} \; d\psi + \int \psi \frac{d\varphi}{ds} \;df
\\
&= -\int \varphi \frac{df}{ds} \; d\psi + \int \psi \frac{d\varphi}{ds}
\frac{df}{ds} \; ds
= -\int \varphi \frac{df}{ds} \; d\psi + \int \psi \frac{df}{ds} \;d\varphi
\\
&=-\int d\biggl( \frac{\psi}{\varphi}\biggr) \; \varphi^2 \frac{df}{ds}
= \int \frac{\psi}{\varphi}\;  \frac{d}{dm} \biggl(\varphi^2 \frac{df}{ds}
\biggr) \; dm
= \int  \psi\varphi \; \frac{d^2 f}{d\tilde{m}\,d\tilde{s}}  \; dm\,.
\end{align*}
Since $\psi$ is arbitrary, the result follows.
\hfill $\square$
\end{proof}
One of the benefits of Proposition \ref{propgen} is that there exists 
a measure $\hat P$ on the path space $C([0,\infty),I)$ such that $X$ is a 
one-dimensional diffusion with scale $\tilde s$ and speed measure $\tilde m$, 
see Rogers \& Williams \cite[V.47]{RW2}. Note that, in the case where $Z$ is a martingale, 
$\hat P= \tilde P$. As a slight abuse of notation, we write again 
$\tilde P$ instead of $\hat P$, even if $Z$ is not a martingale. 
Note that $\RN=\RN_n$ on $\sF_{T_n}$, for every $n$.
%
So what we have to determine is
\begin{quote}
{\sc Question 1:} Under $\RN$, does $\tilde{Z}$ reach zero in finite time?
\end{quote}
If not, then $Z$ is a martingale.

\medskip
From now on, we shall make the simplifying assumption 
\begin{assumption}\label{ass:A}
 $\varphi$ has a continuous density with respect to $m$.
\end{assumption}
Since $\varphi \in \sD$ by assumption, it is automatic that $\varphi$ has a continuous density with respect to $s$, but in general not with respect to $m$. Assumption \ref{ass:A} would hold if both $s$ and $m$ had continuous densities with respect to Lebesgue measure, for example, a situation covering many examples of interest.

\medskip

Next, the $\RN$-local martingale $\tilde{Z}$ can be expressed as
\begin{eqnarray}
\tilde Z_t &=& \exp \biggl(\;-\log \big(\varphi(X_t)\big) - \int_0^t
\varphi{\tilde{\sG}(1/\varphi)}(X_u) \; du
                \;\biggr)
  \label{eq21}\\
                &=& \exp \bigl(\; \tilde M_t - \half \langle\tilde  M \rangle_t\;\bigr)
                \label{eq22}
\end{eqnarray}
for some continuous $\RN$-local martingale $\tilde M$, where the representation \eqref{eq21} follows from the equality $\varphi^{-1} \sG \varphi = - \varphi \tilde{\sG}(1/\varphi)$, an immediate consequence of \eqref{Gtilde_def}.  
If we make the It\^o expansion of $\log \tilde Z$ as given in \eqref{eq21}, we find after some calculations and simplifications that the finite-variation part of $\log \tilde{Z}$ is
\begin{equation}
-\half \tilde{h}(X_t) \; dt ,
\label{eq23}
\end{equation}
where 
\begin{equation*}
\tilde{h} = \frac{d\varphi}{d\tilde{m} } \frac{d\varphi}{ds}
= \frac{1}{\varphi^2}\frac{d\varphi}{d\tilde{m} } \frac{d\varphi}{d\tilde{s}}
=\frac{1}{\varphi^2}\frac{d\varphi}{d{m} } \frac{d\varphi}{d{s}}.
\end{equation*}
  Hence by comparing \eqref{eq22} and \eqref{eq23} we learn that
\begin{equation}
d \langle \tilde{M }\rangle_t = \tilde{h}(X_t)\, dt\,.
\label{dM2}
\end{equation}
In particular, $\tilde h$ is non-negative.
  So under Assumption \ref{ass:A}
it is now clear that to answer Question 1, we have to answer:
\begin{quote}
{\sc Question 2:} Does 
\begin{equation}
\tilde{A}_t :=  \int_0^t \tilde h(X_u) \; du 
\label{Atilde}
\end{equation}
 reach infinity in finite time?
\end{quote}

\begin{remark} When the diffusion is specified as the solution to an SDE,
\eqref{Atilde} appears at  equation (9) in Mijatovi\'{c} \& Urusov \cite{MU2012b}, equation (2.6) in
\cite{MU2012a}.  If $X$
is the solution to an SDE, then
\begin{align*}
s'(x)&=\exp\Big(-2\int_{x_0}^x \frac{\beta(u)}{\sigma(u)^2} du\Big)\\
m'(x)&=\frac{1}{\sigma(x)^2s'(x)}\,,
\end{align*}
so 
\begin{align*}
\tilde h&=\frac{1}{\varphi^2}\frac{d\varphi}{dm } \frac{d\varphi}{ds}
=\left(\frac{\varphi'}{\varphi}\right)^2 \frac{1}{m' s'}
=\left(\frac{\varphi'}{\varphi}\right)^2 \sigma^2
=\left(\log(\varphi)'\right)^2 \sigma^2
=\left(\frac{c}{\sigma}\right)^2 \sigma^2=c^2\,.
\end{align*}
\end{remark} 
%
%
Of course, Question 2 must be answered in the law $\RN$.
If $K\subset \Int$ is any compact set, and $\zeta = \inf\{t: \tilde A_t =
\infty\} < \infty$, then clearly $X$ must exit $K$ before $\zeta$, because the
integrand in \eqref{Atilde} is bounded on $K$,  
by Assumption \ref{ass:A}.
By considering
an increasing sequence of compact $K_n$ increasing to $\Int$, we see that if
$\tilde A$ reaches infinity in finite time, {\em it has to be at a time when $X$
reaches a boundary point of $I$.}

\medskip
To understand this, we look at the diffusion $Y = \tilde{s}(X)$, which is a diffusion in natural scale under $\RN$, taking values in the interval $\tilde{s}(I)$, whose endpoints are $\tilde{a} := \tilde{s}(a) < \tilde{b} := \tilde{s}(b)$. Two cases arise.

\medskip\noindent
{\sc Case 1:} {\em $\tilde{a}$ and $\tilde{b}$ are both infinite}. Since $Y$ is a continuous local martingale, and therefore a time-change of Brownian 
motion, see, e.g., Revuz \& Yor \cite[Chapter V, Theorem 1.7]{RY2004}, $Y$ cannot reach either endpoint  in finite time, so the change-of-measure local martingale $Z$ is a true martingale.

\medskip\noindent
{\sc Case 2:} {\em one at least of $\tilde{a}$ and $\tilde{b}$ is finite.} To fix ideas, let us suppose that $\tilde{a}$ is finite and $\tilde{b}=\infty$ , and see what happens at $\tilde{a}$; the treatment at a finite upper boundary point is analogous.

Firstly we have to ask whether $Y$ reaches the lower boundary point $\tilde a$
in finite time. According to Proposition \ref{prop1} (with $h\equiv 1$, cf 
Remark \ref{rem:prop1} (ii)
), this happens if and only if
\begin{equation}
\int_{a+}\big(  \tilde{s}(x) - \tilde{s}(a) \big) \; \tilde{m}(dx)   < \infty.
\label{reach_zero}
\end{equation}

 If $Y$ does {\em not} reach $\tilde a$ in finite time, then explosion of $\tilde A$ in finite time is clearly impossible. 
 
 However, if $Y$ {\em does} reach $\tilde a$ in finite time, then the additive functional $\tilde{A}$ may explode at that time. In the situation considered by \cite{MU2012b}, where the diffusion $X$ stops at $a$ if it ever gets there, then the criterion for explosion is (according to Proposition \ref{prop1})
 \begin{equation}
  \int_{a+} \big( \tilde{s}(x)-\tilde{s}(a)\big) \, \tilde h(x) \;\tilde{m}(dx) = \infty.
  \label{explode_MU}
  \end{equation} 
On the other hand, if the diffusion $Y$ reflects off $\tilde a$, then explosion
could happen at $H_a+$ even though there was no explosion at $H_a-$, and the
criterion now for $\tilde A$ to explode at $H_a+$ is 
\begin{equation}
\int_{a+}  \, \tilde h(x) \;\tilde{m}(dx) = \infty,
  \label{explode}
\end{equation}
by Proposition \ref{prop1}, part 2. For the applications of interest to us, this is the relevant criterion, as the CIR diffusions we deal with later all reflect off the boundary point.

Notice that condition \eqref{explode} can be equivalently expressed (due to the form \eqref{dM2} of $\tilde h$ and \eqref{tilde_ms}) as 
\begin{equation*}
\int_{a+} \biggl(\frac{d\varphi}{d s}(x)\frac{d\varphi}{d m}(x)\biggr) \;  m(dx) =\infty\,.
\end{equation*}
Thus we see that in order to decide whether the local martingale $Z$ is {\em not} a true martingale, we have to answer the three questions:
\begin{enumerate}
\item Is at least one of the endpoints $\tilde{a}$, $\tilde{b}$ of $\tilde{s}(I)$ finite \; ?
\item If $\tilde{a}$ (say) is finite, does $X$ reach $a$ in finite time (see \eqref{reach_zero})\; ?
\item If so, does $\tilde A$ explode when $X$ reaches $a$ (see \eqref{explode})\;?
\end{enumerate}
To summarise then, we have the following result.


\begin{theorem}\label{thmA}
Let $X$ be a diffusion on the interval $I \subset \R$, $X_0 = x_0 \in I$, $a := \inf I$, $b := \sup I$, with scale function $s$ and speed measure $m$.
We define the change-of-measure local martingale $Z$ by
\begin{equation*}
Z_t = \varphi(X_t)\;\exp \bigg( - \int_0^t \frac{\sG \varphi}{\varphi}(X_u)
\; du \biggr)\,,
\end{equation*}
where $\sG$ is the generator of $X$ and $\varphi$ is strictly positive and $C^1$ with $\varphi(x_0) = 1$.
We let $\RN$ be the probability defined by the change-of-measure local martingale $Z$ via \eqref{com}, defined on the $\sigma$-field $\sF_{T_\infty}$, and we denote the scale function and speed measure of $X$ under $\RN$ by
 $\tilde{s}$, $\tilde{m}$ respectively. These are related to $s$, $m$ as given at \eqref{tilde_ms}. Denote 
 $\tilde a = \tilde{s}(a)$, $\tilde b = \tilde{s}(b)$. 
Assume further Assumption \ref{ass:A}.
 \medskip
 
 If all of the following three conditions are satisfied:
\begin{enumerate}
\item At least one of the endpoints $\tilde{a},\; \tilde{b}$ is finite;
\item At least one of the finite endpoints is reached in finite time (see
\eqref{reach_zero});
\item There is a finite endpoint which is reached in finite time and at which the additive functional $\tilde A$ explodes (see \eqref{explode}  or \eqref{explode_MU}),
\end{enumerate}
then the change-of-measure local martingale $Z$ is not a true martingale. Otherwise, it is.
\end{theorem}


\begin{remark} 
In \c{C}etin \cite{Ce2018} a similar question to ours is discussed, by giving a class of
absolutely continuous measure changes using potentials.  More specifically, we
characterise the positive functions $\varphi$ for which $Z$ is a true
martingale, whereas \cite[Theorem 3.2]{Ce2018} shows that if $\varphi$ is a
potential, then Z is a true martingale.  The key differences are that
\cite{Ce2018} works in an SDE setting and that the statement of 
\cite[Theorem 3.2]{Ce2018} assumes  $X$ has a semimartingale local time, but
the most general one-dimensional diffusion such as we work with does  not have
to be a semimartingale. 
\end{remark}

If $Z$ is a martingale, then the recipe \eqref{com} defines a new measure $\RN$ on the path space under which the canonical process is again a regular diffusion. The law $\RN$ is therefore absolutely continuous with respect to $\RW$, but not in general equivalent. Here are two interesting examples, where the process is given by the SDE recipe \eqref{dXbis} and the change-of-measure local martingale is of the form \eqref{Zsde}.

\medskip

\begin{example}
A canonical example is when $X$ solves \eqref{dXbis} with $\sigma(x) \equiv
1$, $\bb(x) \equiv 0$, on $I=[0,\infty)$ with $X_0=x_0>0$. 
That is, $X$ is of the form $X_t=x_0+W_t^{H_0}$, where $W$ is a
standard Brownian motion and $H_0$ is the time when $X$ hits $\{0\}$. 
We want $c(x) = 1/x$ so that $X$ solves the 
BES(3) SDE
\[
dX_t=\frac{1}{X_t} dt+d\tilde W_t\quad X_0=x_0\,,
\] under $\RN$. In this example, $a=0$, $b=\infty$, $s(x)=x$ and
$m'(x)=1$. From \eqref{logphi} and \eqref{tilde_ms} we find that $\varphi(x) = x/x_0$, and we may take $\tilde{s}(x)
= -x_0^2/x$, $\tilde{m}'(x) = x^2/x_0^2$, and therefore $\tilde{a} = -\infty$,
$\tilde{b} = 0$. 
According to our method we next ask whether the finite boundary point $\tilde
b$ can be reached in finite time. By the integral test \eqref{reach_zero} (in
the analogous form for an upper boundary) the process $X$ approaches $\infty$
(or: $\tilde s(X)$ approaches 0) under $\RN$, but never gets there.

Thus by Theorem \ref{thmA} there is an absolutely continuous change of measure,
taking Wiener measure $P$ to the law $\RN$  of BES(3) started at $x_0$, which
is absolutely continuous with respect to Wiener measure $\RW$. $\RN$ is not
equivalent to $\RW$, since $Z_t$ is not a.s.~positive.

Note that in this example we knew from the outset that $Z$ is a true martingale,
but nevertheless, the application of our recipe is illuminating. 
\end{example}

\begin{example}
An important example for the CIR process \eqref{dvv} followed by the volatility in the Heston model  is the case where under $\RW$
the diffusion $X$ follows \begin{equation}
dX_t = 2 \sqrt{ X_t^+} \, dW_t  + \delta_0 \, dt,\qquad X_0 = x_0>0, 
\label{besq0}
\end{equation}
the squared-Bessel SDE of dimension $\delta_0 >0$. See Chapter XI of Revuz \& Yor \cite{RY2004} for  a definitive account. Suppose that we want to perform a measure change to transform the SDE to
\begin{equation}
dX_t = 2 \sqrt{ X_t^+} \, d\tilde W_t  + \delta_1 \, dt,\qquad X_0 = x_0>0,
\label{besq1}
\end{equation}
where again $\delta_1>0$.
This requires us to add a drift $c(X_t) dt$ to $dW_t$ in \eqref{besq0}, where
\begin{equation*}
c(x) = (\delta_1 - \delta_0)/(2\sqrt{x}).
\end{equation*}
Simple calculations give us
\begin{equation*}
\varphi(x) = x^{(\delta_1-\delta_0)/4},
\end{equation*}
taking $x_0=1$ with no real loss of generality. The scale function $\tilde{s}$ is given by
\begin{equation*}
\tilde{s}'(x) = \exp \biggl( - \int^x \frac{2 \delta_1}{4y} \; dy \biggr)
= \exp(-\half \delta_1 \log x ) = x^{-\delta_1/2},
\end{equation*}
so that (up to irrelevant constants)
\begin{equation*}
\tilde{s}(x) = 
\begin{cases}
x^{(2-\delta_1)/2} &  \text{ if } \delta_1 \ne 2\\
 \log x&  \text{ if } \delta_1 = 2\,.
\end{cases}
\end{equation*}
There are three cases to understand:
\begin{enumerate}
\item $0 < \delta_1 < 2$. Here, $\tilde a=0$ and $\tilde b=\infty$.
The criterion \eqref{reach_zero} shows that $\tilde a$ is reached in finite time, and
the criterion \eqref{explode} requires us to calculate 
\begin{equation*} 
\begin{split}\int_{a+}  \, \tilde h(x) \;\tilde{m}(dx) &= 
\int_{0+} \left(\frac{ \varphi'(x)}{\varphi(x)}\right)^2\frac{1}{\tilde{s}'(x)}
\; dx \\
&= \frac{(\delta_0-\delta_1)^2}{16} \int_{0+ } x^{-2+ \delta_1/2} \; dx 
= \infty,
\end{split}
\end{equation*}
so in this case there is {\em never} an absolutely continuous change of measure which achieves the desired drift, whatever $\delta_0 \neq \delta_1$. 
\item $\delta_1 = 2$. In this case, $\tilde{s}(x) = \log x$, thus $\tilde a=-\infty$ and $\tilde b=\infty$. So the first check of our recipe fails, and there is an absolutely continuous measure change that achieves the desired drift.
\item $\delta_1 >2$. This time, $\tilde{s}(x) = - x^{-(\delta_1-2)/2}$, so
$\tilde b=0$,  $\tilde a=-\infty$. However, the criterion
\eqref{reach_zero} is infinite for approaching $\tilde b$, so $X$ approaches but
never reaches $\infty$ under $\tilde P$, and there is an absolutely continuous
measure change which turns the dynamics of $X$ into \eqref{besq1}.
\end{enumerate}
So to summarize, if we want to use a change of measure to change the dimension of a BESQ($\delta_0$)  to $\delta_1 \neq \delta_0$, this is
\begin{itemize}
\item {\em never} possible if $\delta_1<2$;
\item {\em always} possible if $\delta_1 \geq 2$.
\end{itemize}
\end{example}

\section{Arbitrage opportunities in the Heston model.}\label{S3}
As is well known, the SDE \eqref{dvv} for the Heston volatility has a pathwise-unique strong solution from any non-negative starting point.
The following fact about the strict positivity of a CIR process is  also well-known; see for example G\"oing-Jaeschke \& Yor \cite{GY2003}.

\begin{lemma}\label{lemma:feller-cond}
For the CIR process $\vv$ specified by \eqref{dvv} 
the following are equivalent:
\begin{enumerate}
\item[(i)] 
$\RW[ \forall t\in (0,T]\colon  \vv_t>  0 ] = 1 $ \\[-0.5em]
\item[(ii)] \(2 \kappa \theta \ge \sigma^2 \) (Feller condition)
\end{enumerate}
\end{lemma}

By scaling time in the CIR SDE \eqref{dvv} to convert the volatility $\sigma$
to the canonical value 2 appearing in the BESQ SDE \eqref{besq0}, we see that
the Feller condition is equivalent to the statement that the {\em effective
dimension} of the CIR process is at least 2, 
\begin{equation}
\delta := \frac{4\kappa\theta}{\sigma^2}  \geq 2.
\label{Feller}
\end{equation}

\begin{definition}
A probability measure $\RN$ on $\sF$ is an 
{\em equivalent local martingale measure (ELMM)} if
\begin{enumerate}
\item[(i)] for all $A\in \sF$ one has $\RN[A]=0$ iff $\RW[A]=0$;
\item[(ii)] the process $S$ is a local martingale under $\RN$.
\end{enumerate}
\end{definition}

The following Lemma \ref{lemma:MPOR_var} is a direct consequence 
of standard results about ELLMs in market models, 
 which can be 
found, for example, in Lemma 5.4.2 and Theorem 5.4.3 of Williams \cite{Williams2006}.

\begin{lemma}\label{lemma:MPOR_var}
Let $\RN$ be an ELMM for the generalized Heston model.
Then there exist previsible processes $\gamma, \gamma'$, both locally 
square-integrable, such that 
\begin{enumerate}
\item[(i)] the process $Z$ with $Z_t:= e^{M_t-\frac{1}{2}[M]_t}$  with
\begin{equation*}
M_t:= \int_0^t \gamma_t\; dW_t+\int_0^t \gamma'_t\;dW'_t,\quad
t\in [0,T]\,;
\end{equation*}
is a martingale;
\item[(ii)] $Z_T$ is a density for $\RN$;
\item[(iii)] The integrand $\gamma$ satisfies
\begin{equation}
 \mu(\vv_t)+\sqrt{\vv_t}\,(\rho\gamma_t+\rho'\gamma'_t) =0\; \hbox{\rm for a.e.} \;t\in [0,T]
\label{gamma_condition}
 \end{equation} 
\item[(iv)] $(S_t)_{t\in [0,T]}$ is a local martingale with respect to $\RN$ iff 
$(Z_tS_t)_{t\in [0,T]}$ is a local martingale with respect to $P$.
\end{enumerate}
\end{lemma}


\begin{theorem}\label{th:Heston_arbitrage}
Suppose that the Feller condition \eqref{Feller} fails: $\delta <2$.  Then
\begin{enumerate}
\item the generalized Heston model admits no ELMM if $\mu(0)\neq 0$;
\item the generalized Heston model has an ELMM if
\begin{equation}
\int_{0+} \mu(x)^2 x^{-2 + \delta/2} \; dx < \infty.
\label{condition}
\end{equation}
\end{enumerate}
\end{theorem}

\begin{proof}
\noindent
1.:
We prove this by contradiction. So assume that there does exist an ELMM.
By Lemma \ref{lemma:MPOR_var}  there exists a
martingale $Z$ such that 
$\big( S_tZ_t \big)_{t\in [0,T]}$ is a local martingale and
\[
dZ_t/Z_t=\gamma_t dW_t
+\gamma'_t dW'_t \,, \quad (t \in [0,T])
\]
with previsible locally square-integrable, and therefore a.s.~pathwise square integrable 
processes $\gamma, \, \gamma'$ satisfying \eqref{gamma_condition}.
Using the continuity of $\mu$,  $\mu(0)\ne 0$, and the fact that the 
$\gamma, \, \gamma'$ are square-integrable, this implies 
\begin{equation}
\label{eq:infty1}
\int_0^T \frac{1}{\vv_t}\; dt<\infty \quad\text{$\RW$-a.s.}
\end{equation}
By Lemma \ref{lemma:feller-cond} $\RW[\forall t\in [0,T]\colon \vv_t>0]<1$.
Therefore, if we define  
\[\tau_0:= \inf\{t\ge 0\colon \vv(t)=0\}\wedge T\]
 we have $\RW[\tau_0<T]>0$, and, in particular 
\begin{equation}
\label{eq:nu-tau0}
\RW[\vv_{\tau_0} =0]>0\,.
\end{equation}
On the other hand, by It\^o's formula,
\begin{align}
\label{eq:nu-ito}
\log(\vv_{\tau_0})
&=\log(\vv_0)+\int_0^{\tau_0}\frac{\sigma}{\sqrt{\vv_t}}\; dW_t
+\int_0^{\tau_0}\Big(\frac{2\kappa\theta-\sigma^2}{2\vv_t}-\kappa\Big)\; dt\, .
\end{align}
From \eqref{eq:infty1} we get 
\begin{align*}
\int_0^{\tau_0} \frac{1}{\vv_t}\; dt<\infty \quad\text{$\RW$-a.s.}\,,
\end{align*}
so that both integrals in  \eqref{eq:nu-ito} are finite a.s., and therefore
we obtain that
\(\RW[\log(\vv_{\tau_0})>-\infty]=\RW[\vv_{\tau_0}>0]=1\,.\)
But this contradicts \eqref{eq:nu-tau0}.

\medskip\noindent
 2.: It is to be expected that if there is an ELMM then there will be many, so to prove the second statement we shall identify one. 

We choose to take the change-of-measure martingale to be
\begin{equation}
\frac{dZ_t}{Z_t}  = -\frac{\mu(\vv_t)}{\rho' \sqrt{\vv_t}}  \; dW'_t
:= c(\vv_t) \; dW'_t, \qquad Z_0 =1.
\label{Z2}
\end{equation}
We see that {\em provided $Z$ is a martingale} the drift of $S$ becomes 0, and the dynamics of $\vv$ is unchanged. So we need show that $Z$ is a true martingale, and for this we use Theorem \ref{thm1} and the arguments of Section \ref{S2}. As before at \eqref{tildeZdef}, we define
\begin{equation*}
\tilde{Z_t} := \frac{1}{Z_t} = \exp\biggl(
\int_0^t -c(\vv_s)  \; d\tilde{W'}_s
-\half \int_0^t c(\vv_s)^2 \; ds
\biggr).
\end{equation*}
Here, $d\tilde{W'}_t = dW'_t +c(\vv_t) \, dt$. Noticing that  $\tilde{Z}$ can be written
\begin{equation*}
\tilde Z_t = \exp\big(
B_{A_t} - A_t 
\big)
\end{equation*}
 for some Brownian motion $B$, with $A_t := \int_0^t c(v_s)^2 \; ds$, it is clear that Question 1 from Section \ref{S2}
is now equivalent to 
\begin{center}
{\sc Question 2':} Does $A_t := \int_0^t c(v_s)^2 \; ds$ reach infinity in finite time?
\end{center}
This is a question about the CIR process $v$. The scale function of $v$ is given by
\begin{equation*}
\tilde s'(v) = \exp\biggl( -2 \int_0^v \frac{\kappa(\theta-x)}{\sigma^2 x} \; dx
 \biggr)
= v^{-\delta/2} e^{2\kappa v/\sigma^2}.
\end{equation*}
The scale function $\tilde s$ is therefore finite at 0, since $\delta<2$, and $\vv$ will reach 0 in finite time. The criterion that $A$ does not explode is (see Proposition \ref{prop1}) that
\begin{equation}\label{eq:criterion-v}
\int_{0+}  \, c(x)^2 \;\tilde{m}(dx) =  \int_{0+} c(x)^2 \; \frac{dx}{\sigma(x)^2\tilde{s}'(x)}
\asymp \int_{0+} \mu(x)^2 x^{-2+\delta/2 } \; dx
\end{equation}
should be finite, and this is condition \eqref{condition} (The symbol $\asymp$ 
in \eqref{eq:criterion-v} means that the ratio of the two sides is bounded and bounded away from~0).
\hfill $\square$
\end{proof}

\begin{remark}
Similar calculations as those in the preceding proof of the first statement appear in Guo \cite{G2008}, where it is shown that there is no ELMM if the stock 
price process itself is a CIR process and the Feller condition does not hold.
\end{remark}

\begin{corollary}\label{cor:Heston_classic}
The classical Heston model with constant drift $\mu\ne 0$ does 
not admit an ELMM if the Feller condition is not satisfied.
\end{corollary}

The significance of this result lies in the fact that by the
famous {\em fundamental theorem of asset pricing (FTAP)} 
the non-existence of an ELMM implies the existence of a {\em 
free lunch with vanishing risk}, i.e.~a weak form of arbitrage, see Delbaen \& Schachermayer \cite{DS1994,DS1998}. We give its explicit construction in the Appendix.

Finally, for completeness, we record this little result which tells us what happens in the case when the Feller condition holds.

\begin{theorem}
Suppose that the Feller condition \eqref{Feller} holds: $\delta \geq 2$.  Then there is always an ELMM.
\end{theorem}

\medskip\noindent
\begin{proof}
Recall our standing assumption that $\mu$ is continuous. We will use exactly
the same change-of-measure martingale \eqref{Z2} as we used for the proof of
Statement 2 of Theorem \ref{th:Heston_arbitrage}. Exactly as in that proof, we
need to establish that $A_t := \int_0^t  c(\vv_s)^2 \; ds$ remains finite
for all time. But we have \begin{equation*}
c(v) := - \frac{\mu(v)}{\rho'\sqrt{v}},
\end{equation*}
and since the CIR process remains strictly positive for all $t>0$ by Lemma \ref{lemma:feller-cond}, and does not explode, it follows immediately that if $\vv_0>0$ then $A$ does not explode. It $\vv_0=0$, a separate argument is required, which we leave to the reader.
\hfill $\square$
\end{proof}

\section{Conclusion}\label{con}

We have provided a complete characterization of when the change-of-measure
local martingale that transforms a one-dimensional diffusion to another one
with a different drift is a true martingale. We are able to decide this
question by a simple three-step-algorithm (compare Theorem~\ref{thmA}). This
has practical implications for a generalized Heston model that allows for a
volatility-dependent growth rate: We can show absence of arbitrage given that a
simple integrability condition holds, even when the Feller condition is
violated. This extends the results for the classical Heston model with constant
growth rate different from the riskless rate, for which we have shown that no
ELMM exists and thus arbitrage opportunities are incurred in that case.


\appendix \normalsize

\section{Making an FLVR in the Heston model.}\label{app}
The main result of Delbaen \& Schachermayer \cite{DS1994} is that for a locally bounded semimartingale
the existence of an equivalent  local martingale measure is a condition
equivalent to the absence of a free lunch with vanishing risk. The following lemma follows readily from the definition of  
{\em free lunch with vanishing risk (FLVR)}, see \cite[Definition 2.8]{DS1994}.

\begin{lemma}\label{FLVR}
Suppose that there exists a sequence $f_n := (H_n \cdot S)_\infty$ of admissible terminal wealths with the properties:
\begin{enumerate}
\item the negative parts $f_n^-$ tend uniformly to zero;
\item the $f_n$ tend almost surely to some non-negative limit 
$f_\infty$ which is not almost surely zero.
\end{enumerate}
Then there exists a FLVR.
\end{lemma}

\begin{proof}
We need to construct a sequence $(K_m)$ of admissible strategies 
and a sequence $(g_m)$ of bounded measurable functions such that $(K_m\cdot S)_\infty\ge g_m$
and a measurable function $g_\infty$ which is non-negative, positive with positive probability,
such that $\lim_m \|g_m-g_\infty\|_\infty=0$.

Without loss of generality we may assume that $f_n\ge -n^{-1}$ for all $n$, by passing to
a subsequence if necessary.
We have $f_n\wedge 1\to f_\infty\wedge 1$ a.s. Note that 
$-1\le -n^{-1}\le f_n\wedge 1\le 1$ and $0\le f_\infty\wedge 1\le 1$. 
Let $A:=\{f_\infty\wedge 1>0\}$. By assumption,
$P(A)>0$. By Egorov's theorem, there exists a measurable set $B\subseteq A$ with  $P(B)>0$ and
a subsequence $(n_m)$ such that,
uniformly on $B$,
\[f_{n_m}\wedge 1\to f_\infty\wedge 1 \;(m\to\infty)\,.
\] 

We now define $K_m:=H_{n_m}$, 
\[g_m:=(f_{n_m}\wedge 1)1_B - n_m^{-1}1_{\Omega\setminus B} \text{ and }
g_\infty:= (f_\infty\wedge 1 )1_B\,,\] 
which have the required properties.
\hfill $\square$
\end{proof}

We shall here directly construct a sequence $(f_n)$ with above properties and thus a FLVR, and then it follows from the result of Delbaen \& Schachermayer \cite{DS1994,DS1998} that there is no equivalent $\sigma$-martingale measure, and {\it a fortiori}  no equivalent local martingale measure.

\medskip

To fix ideas, we shall assume with no real loss of generality that $r=0$, and  that $\mu >0$; if $\mu=0$ then we are already in an equivalent local martingale measure and there is nothing interesting to say, and if $\mu<0$ the argument we give carries through by reversing signs in the appropriate places.

\medskip

We firstly reduce the problem to a simpler canonical form, by modifying the SDE for $v$ to 
\begin{equation}
dv_t = \sigma \sqrt{v_t} \, dW_t + \kappa\theta  \, dt.
\label{dv2}
\end{equation}
We can always do this, because if we can construct a FLVR in this setting we
can perform an absolutely-continuous change of measure to change the drift in
\eqref{dv2} into the original drift in \eqref{dvv}, and null events (and
therefore an  FLVR) will not be changed by this (of course, this will change
drift in \eqref{dS1}, but an equivalent change of measure to $W'$ can be
applied to reverse this).  
Once we have done this, we have that $v$ is a BESQ
process, or at least, a BESQ process run at a constant speed which may not be
1. Again, we change nothing that matters by rescaling the speed so that we are
looking at an actual BESQ process
\begin{equation}
dv_t = 2 \sqrt{v_t} \, dW_t + \delta \, dt
\label{besq}
\end{equation}
where we have the correspondence $\delta = 4 \kappa\theta/\sigma^2$. Thus the Feller condition \eqref{Feller} is the statement that $\delta <2$, the familiar condition in terms of the dimension $\delta$ of the BESQ process that the process hits 0. For more background on BESQ processes, we refer to Revuz \& Yor \cite{RY2004}.

\medskip

Looking at \eqref{dS1}, it is rather obvious what the idea of the construction should be: we need to go into the asset when $v$ is very small, because at such times the positive drift $\mu$ will dominate the tiny variance. Ideally, we could just hold the asset at the times when $v$ is {\em equal} to zero, because then the martingale part of the gains-from-trade process would vanish and we would just get the drift contribution, but this does not work because the Lebesgue measure of the set of times when $v=0$ is zero; see, for example, Proposition 1.5 on page 412 of \cite{RY2004}.  So the next attempt is to hold the asset only at times when $v_t < \varepsilon$ for some very small $\varepsilon>0$, which we hope will be an approximate arbitrage. As we shall see, this leads us to an FLVR.

\medskip

But in order to do this, we have to be able to do some calculations on BESQ processes, which turn out to be easier in terms of the scale and speed representation of $v$ in terms of a standard Brownian motion. The scale function of $v$ is easily verified to be
\begin{equation*}
s(x) \propto x^{1-\delta/2}.
\end{equation*}
If we then consider the diffusion in natural scale $Y_t = s(v_t) = v_t^{1-\delta/2}$, and apply It\^o's formula (as the scale function is not $C^2$, this is only valid in the region $(0,\infty)$ where $s$ is $C^2$) we find that
\begin{equation}
dY_t = (2-\delta) v_t^{(1-\delta)/2} \; dW_t = (2-\delta) Y_t^{(1-\delta)/(2-\delta)} \; dW_t,
\label{dYt}
\end{equation}
at least while $Y$ is strictly positive. Clearly \eqref{dYt} cannot hold for all time, otherwise $Y$ would be a non-negative local martingale, and would have to stick at 0 once it reaches 0. Of course, this does not happen, and this is because of a local time effect at zero - see Rogers \& Williams \cite{RW2}, V.48.6, for more details.

But \eqref{dYt} tells us that away from 0 the speed measure for $Y$  is
\begin{equation*}
m(dy) = \frac{ dy}{(2-\delta)^2} \; |y|^{2(\delta-1)/(2-\delta)} \;,
\end{equation*}
and the speed measure does not charge 0 because $Y$ spends no time there. So we
may create a weak solution  to \eqref{besq} starting from a standard Brownian
motion $B$ with local time process $\{\localtime(t,x): x \in \R, t \geq 0\}$ by the
recipe 
\begin{eqnarray*}
\pchaf_t &:=& \int \localtime(t,a) \; m(da)
\\
&=& \int_0^t \frac{|B_u|^{2(\delta-1)/(2-\delta)}}{(2-\delta)^2}\; du,
\\
\tau_t &=& \inf\{u: \pchaf_u > t \},
\\
Y_t &=& |B_{\tau_t}|,
\\
v_t &=& Y_t^{2/(2-\delta)} \; ;
\end{eqnarray*}
for more details, see V.47, V.48 in \cite{RW2}.

\medskip

The idea now is to make a portfolio $\varphi(Y_t) /S_t$ so that the gains-from-trade process becomes
\begin{equation}
G_t  := \int_0^t \varphi(Y_u) \, (\sqrt{v_u} \; d\hat{W} + \mu \, du) ,
\label{GFT}
\end{equation}
where $d\hat{W} = \rho dW + \rho' dW'$; see \eqref{dS1}. 

We shall do this in such a way that the local martingale term in \eqref{GFT} is negligible, and the Lebesgue term is not. To explain, when we look at the Lebesgue integral in $G_t$ we see $\mu$ times
\begin{eqnarray*}
\int_0^t \varphi(Y_u) \; du &=& \int_0^t \varphi(|B_{\tau_s}|) \; ds
= \int_0^{\tau_t}  \varphi(|B_u|) \; d\pchaf_u
\\
&=&\int_0^{\tau_t}  \varphi(|B_u|) \;
\frac{|B_u|^{2(\delta-1)/(2-\delta)}}{(2-\delta)^2}\; du.
\end{eqnarray*}
If we now choose $\varepsilon>0$ and define
\begin{equation}
\varphi(x) = (2\varepsilon)^{-1} I_{[0,\varepsilon]}(x)
\cdot x^{ 2(1-\delta)/(2-\delta)} (2-\delta)^2,
\label{phidef}
\end{equation}
we find that
\begin{eqnarray}
\int_0^t \varphi(Y_u) \; du
&=& (2\varepsilon)^{-1} \int_0^{\tau_t} I_{[0,\varepsilon]}(|B_u|) \; du
= \frac{1}{2\varepsilon}\int_{-\varepsilon}^{\varepsilon} \localtime(\tau_t,x) dx
\nonumber
\\
&\rightarrow& \localtime(\tau_t,0) \qquad (\varepsilon \downarrow 0).
\label{Llim}
\end{eqnarray}
The quadratic variation of the martingale part of $G$ is
\begin{eqnarray}
\langle G \rangle_t &=& \int_0^t \varphi(Y_u)^2 v_u \; du
=\int_0^t \varphi(Y_u)^2 \, Y_u^{2/(2-\delta)} \; du
\nonumber
\\
&=& \int_0^{\tau_t} \varphi(|B_u|)^2 |B_u|^{2/(2-\delta)} \; d\pchaf_u
\nonumber
\\
&=& \frac{(2-\delta)^2}{4 \varepsilon^2} \int_0^{\tau_t}
I_{[0,\varepsilon]}(|B_u|) |B_u|^{2} \; du
\nonumber
\\
&=& \frac{(2-\delta)^2}{4 \varepsilon^2} \int_{-\varepsilon}^{\varepsilon}
\localtime(\tau_t,x)\;  |x|^{2}  \; dx
\label{G0}
\\
&\sim& \frac{(2-\delta)^2}{6 }\; \localtime(\tau_t,0) \, \varepsilon\qquad(\varepsilon\downarrow 0).
\nonumber
\end{eqnarray}

From \eqref{G0} therefore
\begin{align*}
\lim_{\epsilon \downarrow 0}\left< G \right>_t& 
 = 0\quad\text{a.s.}\\
\label{eq:GFT_expectation}
\text{and}\quad
E\left[\left< G \right>_{\pchaf(t)}\right]
&=O(\varepsilon)\,.
\end{align*}

Equations \eqref{Llim} and \eqref{G0} are the main parts of what we need, all that remains is to put the bits together.

\medskip

So we let $M_t$ denote the local martingale part of $G_t$, fix some positive time horizon $T$, and construct the FLVR. For this we consider a sequence $\varepsilon=2^{-n}$ of values of $\varepsilon$, and consider the portfolios $\varphi$ given by \eqref{phidef} for the different values of $\varepsilon$. We are only going to use this portfolio until the stopping time which is the minimum of $t=\pchaf_T$ and 
\begin{equation*}
\theta_n :=\inf\{t: |M_t| > n^{-1} \} ,
\end{equation*}
 after which everything stops. 
Now we have (with $M^*_t := \sup \{ |M_u|: u \leq t\}$)
\begin{eqnarray*}
P\big[|M^*_{\pchaf_T \wedge\theta_n}| > n^{-1} \big] &\leq& n^2E\big[ (M^*_{\pchaf_T \wedge\theta_n})^2 \big]
\\
&\leq& 4 n^2 E[ M_{\pchaf_T \wedge\theta_n}^2 ],
\end{eqnarray*}
by Doob's submartingale maximal inequality, and in view of \eqref{G0}  we have the bound
\begin{equation*}
P[M^*_{{\theta_n}} > n^{-1} ] \leq C n^2 2^{-n}
\end{equation*}
for some finite constant $C$. Hence by Borel-Cantelli, for all but finitely
many $n$ we have $M^*_{\theta_n} \leq n^{-1}$ and therefore $\theta_n >
\pchaf_T$. The negative part of $G_{\pchaf_T \wedge \theta_n}$ is no more than
$n^{-1}$, and as we let $n \rightarrow \infty$ the terminal value $G_{\pchaf_T
\wedge \theta_n}$ converges to $\mu \localtime(T,0)$, which is of course
non-negative, and positive with positive probability.

\medskip
The FLVR is constructed.

\section*{Conflict of interest}
The authors declare that they have no conflict of interest.

\section*{Acknowledgments} 
Sincere thanks is due to Ralf Korn for many fruitful discussions in the early stages of this project. We also appreciate helpful comments of David Criens and an anonymous referee. Moreover, we wish to thank Martin Schweizer for a very careful reading of this manuscript.

S. Desmettre und G. Leobacher are supported by the Austrian Science Fund (FWF) projects  F5507-N26 and F5508-N26, which are part of the Special Research Program \textit{Quasi-Monte Carlo Methods: Theory and Applications}. S. Desmettre moreover appreciates support by the DFG-Research Training Group 1932 \textit{Stochastic Models for Innovations in the Engineering Sciences}.


\end{document}